\documentstyle[preprint,aps,floats]{revtex}
\input epsf.tex
\def\DESepsf(#1 width #2){\epsfxsize=#2 \epsfbox{#1}}
\begin{document}
\preprint{\vbox{\hbox{}}}
\draft
\title{
$\bar B^0 \to \pi^+ X$ in the Standard Model }
\author{X.-G. He$^1$, C.-P. Kao$^1$, J.P. Ma$^2$ and S. Pakvasa$^3$}
\address{
$^1$Department of Physics, National Taiwan University, Taipei\\
$^2$Institute of Theoretical Physics, Academia Sinica, Beijing\\
and\\ $^3$Department of Physics and Astronomy, University of
Hawaii, Honolulu, Hawaii, HI 96822 }
\date{June, 2002}
\maketitle
\begin{abstract}
 In this paper we investigate the possibility of
studying $B\to \pi$ form factor using the semi-inclusive decays
$\bar B^0 \to \pi^+ + X_q$. In general $B\to PX$ semi-inclusive
decays involve several hadronic parameters. But for $\bar B^0 \to
\pi^+ X_q$ decays we find that in the factorization approximation,
the only unknown hadronic parameters are the form factors $F^{B\to
\pi}_{0,1}$. Therefore these form factors can be studied in $\bar
B^0 \to \pi^+ X_q$ decays. Using theoretical model calculations
for the form factors the branching ratios for $\bar B^0 \to \pi^+
X_d(\Delta S = 0) $ and $\bar B^0 \to \pi^+ X_s (\Delta S = -1)$,
with the cut $E_{\pi} > 2.1$ GeV, are estimated to be in the ranges
of $(3.1\sim 4.9) \times
10^{-5}(F^{B\to \pi}_1(0)/0.33)^2$ and 
$(2.5\sim 4.2)\times 10^{-5}(F_1^{B\to \pi}(0)/0.33)^2$, 
respectively, depending 
on the value of $\gamma$. The combined branching ratio
for $\bar B^0 \to \pi^+ (X_d+ X_s)$ is about $7.4\times 10^{-5}
(F^{B\to \pi}_1(0)/0.33)^2$ and 
is insensitive to $\gamma$. We also discuss CP asymmetries in 
these decay modes.
\end{abstract}

\pacs{}

\preprint{\vbox{\hbox{}}}

\section{Introduction}

There have been considerable efforts to understand the properties
of $B$ decays. Recently several rare charmless hadronic  $B$
decays have been measured. These decays being rare are sensitive
to CP violating parameter $\gamma$ in the Standard Model
(SM) and also new physics\cite{1,2,3,4,5,6}. While most of the
studies have concentrated on the exclusive $B$ decay modes, there
are also some studies for semi-inclusive decays\cite{3,4,5}, for
example $B\to \eta' X$ have been studied in details
experimentally\cite{7}. At present there are several other multi-body rare
$B$ decays have been measured, such as $B\to \pi\pi K, K K K$\cite{8}. 
It can be expected that more $B\to P X$ decay modes, with
$P$ being a light meson, will be experimentally studied.
Theoretically at the quark level the effective Hamiltonian for $B$
decays is well understood in the SM. The quark-hadron duality
allows one to have a good understanding for inclusive hadronic
decays. But for exclusive and semi-inclusive decays there are more
uncertainties. The major uncertainties come from our poor
understanding of the long distance strong interaction dynamics.

In rare charmless hadronic exclusive two body B decays of the type $B\to PP$,
the operators which induce such decays in the SM,
to the lowest order, are four quark operators $O_i$. In the factorization
approximation, the four quark operators are factorized into bi-quark operators

\begin{eqnarray}
<P_1 P_1|O_i|B> &=& <P_1|j_1|0> <P_2|j_2|B> + <P_1|j'_1|0>
<P_2|j'_2|B> + (1 \to 2, 2\to 1)\nonumber\\ &+& <P_1
P_2|j_1|0><0|j_2|B> + <P_1 P_2|j'_1|0><0|j'_2|B>,
\end{eqnarray}
where $j'_1\times j'_2$ is the Fierz transformed form of
$j_1\times j_2$. The last two terms are referred as the
annihilation contributions which are usually assumed to be small
and are neglected. The bi-quark operators $<P_i|j_1^{(')}|0>$ and
$<P_i|j_2^{(')}|B>$ can be related to $P$ meson decay constants or
$B\to P$ transition form factors. Several of the decay constants,
such as $f_\pi$ and $f_K$ have been measured to good accuracy, but
less is known about form factors. Present experimental measurements
are consistent with model calculations. To have better
determinations of CP violating parameter and possible new physics,
more accurate determination of the form factors is necessary.

Some theoretical studies for semi-inclusive charmless hadronic
decay modes $B\to P X$ have been carried out before\cite{3,4,5}. In
the factorization approximation, the decay amplitude contains
several terms,

\begin{eqnarray}
A(B\to P X) &=& <X|j_1|0><P|j_2|B> + <P|j_1|0><X|j_2|B>\nonumber\\
&+& <PX|j_1|0><0|j_2|B> + (\mbox{Fierz transformed terms}).
\label{facto}
\end{eqnarray}
Compared with exclusive decays, semi-inclusive decays involve fewer
unknown hadronic parameters, therefore semi-inclusive decays may
be better understood from theoretical point of view.
Experimentally semi-inclusive decays may be more difficult to
study because background. In order to make sure that the observed
events are from rare charmless B decays, and other processes, such
as $B\to D(D^*) X' \to P X''$, do not contaminate the direct rare
charmless hadronic $B\to P X$ decays, one needs to make a cut on
the $P$ energy. It has been shown that with a cut of $E_P
> 2.1$ GeV, most of the unwanted events can be eliminated\cite{3}. The
resulting events will have a small invariant mass $M_{X}^2$. With
the cut $E_{P}> 2.1$ GeV, $M^2_{X} < 5.7$ GeV$^2$. Rare charmless
hadronic B semi-inclusive decays can be studied experimentally and
useful information can be obtained. In this paper we study a class
of the semi-inclusive decay $\bar B^0\to \pi^+ X_q$ with the
emphasis on the possibility of using this decays to determine the
form factor $F_{0,1}^{B\to \pi}$.

\section{Decay amplitudes for $\bar B^0 \to \pi^+ X_q$}

From Eq. (\ref{facto}) we see that in general there are three
types of terms in the factorization approximation for the
semi-inclusive decays of the type $B \to P X$.  Each of the terms
involves different hadronic parameters, with the first, the second
and the third terms being proportional to $B$ to $P$ transition
form factor from $<P|j_2|B>$, the $P$ decay constant from
$<P|j_1|0>$, and some other parameters from $ <PX|j_1|0>$ and
$<0|j_2|B>$, respectively. If all three terms in Eq.~(\ref{facto})
contribute with the same order of magnitude to a process, the
accumulated uncertainties will be substantial due to uncertainties
in all the hadronic parameters involved, especially in the form
factors.

If  one or two terms in Eq. (2) can be eliminated, one can have a
better understanding of the processes involved. Indeed this can be
achieved by an appropriate choice of the initial meson $B$ and the
final meson $P$. For example in $\bar B^0\to (K^- X), (\pi^- X)$,
the term proportional to the form factors does not appear\cite{4}, and
allow a good theoretical prediction. We find that there is only
one possible choice for $P$ where the second term is eliminated.
These are the decays $\bar B^0 \to \pi^+ X_q$. Here $X_d$ and
$X_s$ indicate the states having $\Delta S = 0$ and $-1$,
respectively. These decays are directly related to the form
factors $F_{0,1}^{B\to \pi}$. Therefore $\bar B^0 \to \pi^+ X_q$
can be used to study these form factors and to test model
calculations. The form factors can also be studied in
semi-leptonic $B\to l\bar \nu_l \pi$ decays, probably with a
better accuracy compared with that determined from $\bar B^0\to
\pi^+ X_q$. However, the final states are different, one in the
leptonic environment and the other in hadronic environment. They
are complementary to each other.

In $\bar B^0 \to \pi^+ X_q$, the bi-quark operators can only be in 
the forms:
$j_1 = \bar q \Gamma_1 u$ and $j_2=\bar u \Gamma_2 b$ and therefore

\begin{eqnarray}
A(\bar B^0 \to \pi^+ X_q) &=& <X_q|j_1|0><\pi^+ |j_2|\bar B^0> +
<X_q \pi^+|j'_1|0><0|j'_2|\bar B^0>. \label{xfact}
\end{eqnarray}
The second term being annihilation type is sub-leading. We will
neglect its contribution. Note that for $q=s$, the annihilation
term is automatically zero.

We would like to point out that $\bar B^0 \to \pi^+X_q$ is a
many-body decay. It is different from two-body decays. There are, in fact,
more ways of factorization for a many-body decay, such
as$<X_1|j_1|0><X_1' \pi^+|j_2|\bar B^0>$ and $<X_2
\pi^+|j'_1|0><X_2'|j'_2|\bar B^0>$ and , with $X_q = X_1 +X_1' =
X_2 +X_2'$. The two terms in Eq.~(\ref{xfact}) correspond to the
cases: $<X_1'| = <0|$, $<X_2'| = <0|$, respectively. For $\bar
B^0\to \pi^+ X_q$ with a cut $E_\pi > 2.1$ GeV, the final state
$X_q$ has a small invariant mass. This is a quasi-two-body decay,
with $\pi^+$ and $X_q$ moving rapidly apart in opposite
directions. The probability of forming the final state $<X_1'
\pi^+|$ with $<X_1'|\neq <0|$ is less than the probability of
forming the simple final state $<\pi^+|$. The contribution of the
configuration $<X_2\pi^+|j'_1|0><X'_2 |j'_2|B>$ is dominated by
$<X_q\pi^+|j'_1|0><0|j'_2|\bar B^0>$ if it is not zero. The cases
with $|X_1'>$ and $|X_2'>$ not equal to $|0>$ are also higher
order in $\alpha_s$ and therefore suppressed.

We are now ready to present the detailed calculations. The
effective Hamiltonian for rare charmless hadronic $B$ decays at
the quark level is given by

\begin{eqnarray}
H_{eff} &=&{G_F\over \sqrt{2}} \left \{ V_{ub} V^*_{uq}(c_1 O_1
+c_2 O_2) - \sum_{i=u,c,t}\sum_{n=3}^{10} V_{ib}V_{iq}^*c_n^i O_n
\right \}. \label{hamiltonian}
\end{eqnarray}
Here $O_n$ are quark and gluon operators and are given by

\begin{eqnarray}
&&O_1 = (\bar s_i u_j)_{V-A} (\bar u_j b_i)_{V-A},\;\;
O_2 = (\bar s_i u_i)_{V-A} (\bar u_j b_j)_{V-A},\nonumber\\
&&O_{3(5)} = (\bar s_i b_i)_{V-A}\sum_{q'}
(\bar q^\prime_j q^\prime_j)_{V-(+)A},\;\;
O_{4(6)} = (\bar s_i b_j)_{V-A}\sum_{q'}
(\bar q^\prime_j q^\prime_i)_{V-(+)A},\nonumber\\
&&O_{7(9)} = {3\over 2}(\bar s_i b_i)_{V-A}\sum_{q'}
e_{q^\prime}(\bar q^\prime_j q^\prime_j)_{V+(-)A},\;\;
O_{8(10)} ={3\over 2} (\bar s_i b_j)_{V-A}\sum_{q'}
e_{q^\prime}(\bar q^\prime_j q^\prime_i)_{V+(-)A},
\end{eqnarray}
where $(V\pm A)(V\pm A) =\gamma^\mu(1\pm\gamma_5) \gamma_\mu(1\pm
\gamma_5)$, $q^\prime = u,d,s,c,b$, $e_{q^\prime}$ is the electric
charge number of the $q^\prime$ quark, and $i$ and $j$ are color
indices.

The Wilson coefficients $c_n^i$ have been calculated in different
schemes\cite{9,10}. In this paper we will use consistently the scheme
independent NDR results. The values of $c_n$ at $\mu \approx m_b$
with the next-to-leading order (NLO) QCD corrections are given
by\cite{10}

\begin{eqnarray}
&&c_1 = -0.307,\;\;c_2 = 1.147,\;\;c^t_3=0.017,\;\;c^t_4 = -0.037,\;\;
c^t_5=0.010,\;\;c^t_6 = -0.045,\nonumber\\
&&c^t_7= -0.0017\alpha_{em},\;\;c^t_8=0.052\alpha_{em},\;\;
c^t_9=-1.37\alpha_{em},\;\;c^t_{10}=-0.282\alpha_{em},\;\;\nonumber\\
&&c_{3,5}^{u,c} = - {1\over N_c}c^{u,c}_{4,6} = {1\over N_c} P^{u,c}_{s},\;\;
c^{u,c}_{7,9} = P^{u,c}_e,\;\;c^{u,c}_{8,10} = 0,
\end{eqnarray}
where $N_c = 3$ is the number of colors and $\alpha_{em}=1/137 $ is the
electromagnetic fine structure constant. The functions $P^i_{s,e}$ are given by

\begin{eqnarray}
&&P^i_s = {\alpha_s \over 8 \pi} c_2 ({10\over 8} + G(m_I, \mu, k^2)),\;\;
P^i_e = {\alpha_{em}\over 9\pi} (N_c c_1 + c_2)({10\over 9} + G(m_i, \mu, k^2)),
\nonumber\\
&&G(m, \mu, k^2) = 4 \int^1_0 x(1-x)\ln{m^2-x(1-x)k^2\over \mu^2} dx.
\end{eqnarray}

We obtain the decay amplitudes in the factorization approximation as

\begin{eqnarray}
A(\bar B^0 \to \pi^+ X_q) &=& [\alpha_q \bar q \gamma_\mu
(1-\gamma_5) u  + \beta_q \bar q \gamma_\mu (1+\gamma_5) u
][F_1(q^2) (p^\mu_B + p^\mu_\pi) + (F_0(q^2)\nonumber\\
 &-&
F_1(q^2)) {m^2_B -m^2_\pi\over q^2} q^\mu + \gamma_q \bar q
(1+\gamma_5) u {m^2_B - m^2_\pi\over m_b - m_u} F_0(q^2)],
\end{eqnarray}
where $q=p_B - p_\pi$, and

\begin{eqnarray}
&&\alpha_q = {G_F\over \sqrt{2}}[V_{ub}V_{uq}^*({1\over N_c}c_1 +
c_2 + {1\over N_c}c^{tu}_3 + c^{tu}_4 + {1\over N_c} c_9^{tu} +
c_{10}^{tu})\nonumber\\&&\;\;\;\;\;+ V_{cb}V_{cq}^*( {1\over
N_c}c^{tc}_3 + c^{tc}_4 + {1\over N_c} c_9^{tc} +
c_{10}^{tc})],\nonumber\\ &&\beta_q = {G_F\over
\sqrt{2}}[V_{ub}V_{uq}^*(c_6^{tu} + c_{8}^{tu}) + V_{cb}V_{cq}^*(
c_6^{tc} + c_{8}^{tc})],\nonumber\\ &&\gamma_q = {G_F\over
\sqrt{2}}[V_{ub}V_{uq}^*(c_5^{tu} + c_{7}^{tu}) + V_{cb}V_{cq}^*(
c_5^{tc} + c_{7}^{tc})](-{2\over N_c}),
\end{eqnarray}
where $c^{ij} = c^i-c^j$. The above coefficients depend on the
momentum exchange $k^2$. In the heavy b quark limit, $k^2 = m^2_B
(1- 2E_q/m_B)$.

\section{Branching ratios for $\bar B^0 \to \pi^+ X_q$}

From the decay amplitudes obtained in the previous section, we obtain the
differential branching ratio,

\begin{eqnarray}
{d\Gamma\over dx dy} = {m^5_B \over 16 \pi^3}
[(|\alpha_q|^2+|\beta_q|^2) F^2_1 (1-x)(x+y-1) + {1\over 4}
|\gamma_q|^2 F^2_0 (1-x)],
\end{eqnarray}
where $y = 2E_q/m_B$ and $x= 2E_\pi/m_B$.
The physical integration intervals are: $0< x<1$ and $1-x < y <1$.
The branching ratios with appropriate cut on the $\pi^+$ energy
$E_\pi > E_{cut}$ are given by

\begin{eqnarray}
\Gamma(E_\pi > E_{cut}) = \int^1_{2E_{cut}/m_B} dx\int^1_{1-x} dy
{d\Gamma\over dx dy}.
\end{eqnarray}

There are several calculations of the form factors with the value
to be typically in the range of $0.3 \sim 0.4$\cite{11,12}. We will use
$F_{0,1}^{B\to \pi}(0) = 0.33$ 
for illustration. 
For $E_\pi >
2.1$ GeV, the dependence on $q^2$ is small, and we will use a
single pole form as an approximation. For the KM matrix elements
we will use the following independent variables, $V_{us} =
\lambda$, $V_{ub} = |V_{ub}|exp(-i\gamma)$, $V_{cb} = A\lambda^2
$, with $\lambda = 0.2196$, $A = 0.835$, $|V_{ub}| =0.08
|V_{cb}|$. The CP violating phase $\gamma$ it treated as a free
parameter. The results for the branching ratios are shown in
Figure 1. We see that the branching ratios for $\bar B^0\to \pi^+
X_d$ and $\bar B^0 \to \pi^+ X_s$ in the ranges of $(3.1\sim
4.9)\times 10^{-5}$ and $(2.5\sim 4.2) \times 10^{-5}$ respectively.
These can be reached at B factories.

\begin{figure}[htb]
\centerline{ \DESepsf(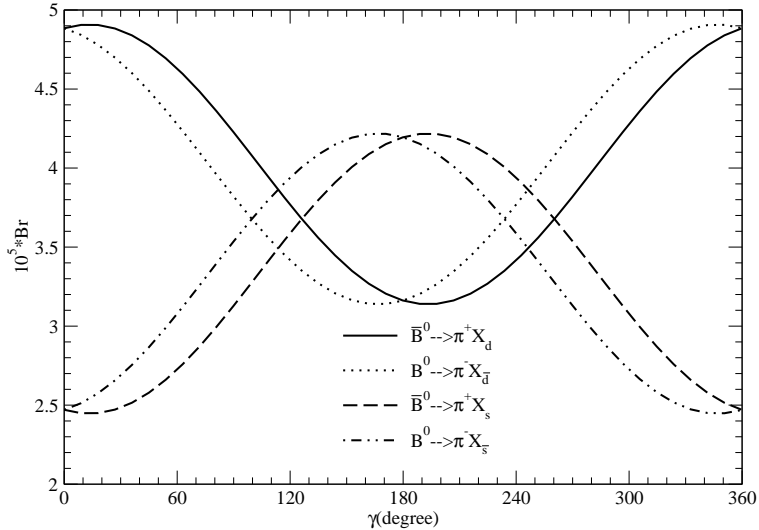 width 10cm)}
\smallskip
\caption {
The branching ratios as a function of $\gamma$.}
\end{figure}

We now would like to point out  an interesting prediction in the
SM regarding rate differences. Due to the unitarity property of
the KM matrix elements, $Im(V_{ub}V_{ud}^* V_{cd}V_{cb}^*) =
-Im(V_{ub}V_{us}^* V_{cs}V_{cb}^*)$, the rate differences,
$\Delta_d =  \Gamma(\bar B^0 \to \pi^+ X_d) - \bar \Gamma(B^0 \to
\pi^+ \bar X_d)$ and the corresponding rate difference $\Delta_s$
have the same magnitude but opposite sign. When the final states
$X_d$ and $X_s$ are not distinguished, one would get zero for
the asymmetry, A,

\begin{eqnarray}
A = { \Gamma(\bar B^0 \to \pi^+ (X_d+X_s)) - \bar \Gamma (B^0 \to \pi^-
(\bar X_d + \bar X_s)) \over \Gamma(\bar B^0 \to \pi^+ (X_d+X_s)) +
\bar \Gamma (B^0 \to \pi^- (\bar X_d + \bar X_s))}.
\end{eqnarray}
This can provide a test for the Standard Model.

We also studied CP asymmetries, with the energy cut $E_\pi > 2.1$
GeV, defined by

\begin{eqnarray}
A_{CP} = {\Gamma(\bar B^0 \to \pi^+ X_q) - \bar \Gamma(B^0 \to
\pi^- \bar X_q) \over  \Gamma(\bar B^0 \to \pi^+ X_q) + \bar \Gamma(B^0 \to
\pi^- \bar X_q)}.
\end{eqnarray}
The results  for $A_{CP}$ are shown in Figure 2. Since $\Delta_d =
-\Delta_s$, because $Br(\bar B^0\to \pi^+ X_d)$ has a larger 
value than $Br(\bar B^0 \to \pi^+ X_s)$ it is easy to 
understand why the asymmetry $A_{CP}$ for
$\bar B^0 \to \pi^+ X_s$ is larger than $\bar B^0 \to \pi^+ X_d$.
The asymmetry in $\bar B^0 \to \pi^+ X_d (X_s)$ can be as large as 5\% 
(6\%).

\begin{figure}[htb]
\centerline{ \DESepsf(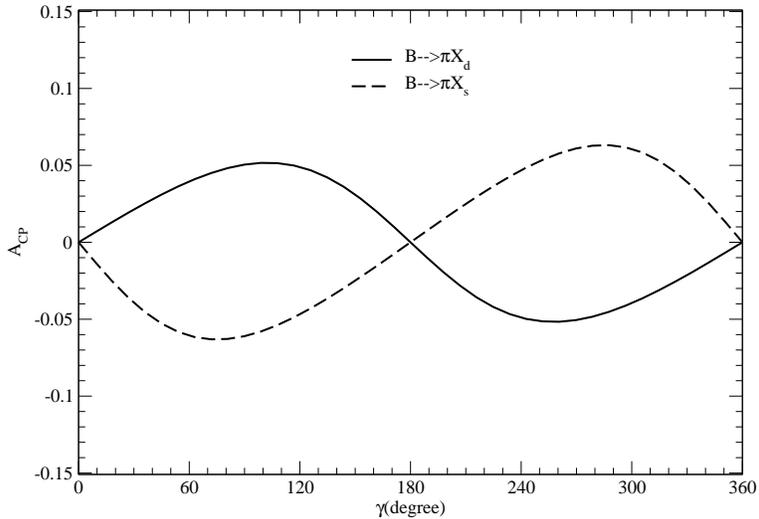 width 10cm)}
\smallskip
\caption {
The CP asymmetries as a function of $\gamma$.}
\end{figure}

\section{Discussions and Conclusions}

From discussions in the previous sections, it can be seen that the
measurements of the branching ratios for $\bar B^0 \to \pi^+ X_q$
can obtain information about the form factors $F^{B\to
\pi}_{1,0}$. If the form factor is known, the
branching ratios can be predicted. The numerical values are
obtained using $F_1(0) = F_0(0) = 0.33$. In general $F_1(q^2)$
and $F_0(q^2)$ have different dependence on $q^2$, one would
expect that several hadronic parameters are needed to specify the
details.
However since 
$F_1(0) = F_0(0)$ for small $q^2$, such as
the case with $E_\pi > 2.1$ GeV, $F^{B\to \pi}_1(q^2) \approx F^{B\to \pi}_0(q^2) \approx F^{B\to \pi}_1(0)$, the branching ratios to a good
approximation are proportional to $F^2_1(0)$. The branching ratios
obtained can be normalized as, $Br(F_1)  = Br(F_1(0) = 0.33))
(F_1(0)/ 0.33)^2$. We have checked numerically with another set of
realistic form factors having different $q^2$ dependence in Ref.\cite{12}. We
indeed find that in the kinematic region we are interested, the
results do not change very much.

We have argued previously that experimental measurements for $\bar
B^0 \to \pi^+ X_d$, although difficult, can be carried out with an
appropriate cut on the energy of the the pion $E_\pi$. Combining
the measurements with $X_q = X_d$ and $X_q = X_s$ can further
enhance the statistic significance. We find that 
the combined branching ratio for $\bar B^0 \to \pi^+ (X_d + X_s)$
is 
$\sim 7.4\times 10^{-5}(F^{B\to \pi}_1(0)/0.33)^2$ which is insensitive to the
phase angle $\gamma$. 
This implies that even without a good determination of $\gamma$, one
can have useful information about the form factors. 
The combined branching ratio also makes the task 
easier in
that one does not need to distinguish whether the final states
$X_q$ contains different strange numbers $-1$ or 0. 
To ensure that
the $\pi^+$ is from $\bar B^0 \to \pi^+ X_d$, and
not from $B^0 \to \pi^+ \bar X_u$, tagging on $\bar B^0$
or $B^0$ is necessary this can be carried out at B
factories. Measurements of $\bar B^0 \to \pi^+ X_q$ are therefore possible.

There are several uncertainties involved. Some of the main
uncertainties are from the KM matrix elements and the phase
$\gamma$. 
At present the best fit value for $\gamma$ is around
$60^0$\cite{2}. If one uses the modes $\bar B^0 \to \pi^+ X_d$ and 
$\bar B^0 \to \pi^+ X_s$ individually, one need a good knowledge of 
$\gamma$ to obtain precise information about the form factors. However, 
since the combined branching ratio has a very weak dependence on $\gamma$, 
using the combined ratio to obtain information on the form factors does 
not require a precise determination of $\gamma$. 
We studied the sensitivity of the branching ratios
on the magnitudes of 
the KM matrix elements. The largest one may come from the magnitude of 
$V_{ub}$. The branching ratio of $\bar B^0 \to \pi^+ X_d$ is almost
proportional to $|V_{ub}|^2$ which can be easily rescaled. The
branching ratio of $\bar B^0 \to \pi^+ X_s$ is less sensitive to $|V_{ub}|$.  
For an accurate determination of the form factors, a good 
knowledge of the magnitude of the KM matrix
elements, especially $V_{ub}$ is important. 

In this paper we have studied a class of semi-inclusive charmless
hadronic B decays $\bar B^0 \to \pi^+ + X_q$. We find that for
these decays, in the factorization approximation, the only unknown
hadronic parameters are the form factors $F^{B\to \pi}_{0,1}$.
Accurate measurement of these decays can provide important
information about form factors. Using theoretical model
calculations for the form factors the branching ratios for $\bar
B^0 \to \pi^+ X_d(\Delta S = 0) $ and $\bar B^0 \to \pi^+ X_s
(\Delta S = -1)$, with the cut $E_{\pi} > 2.1$ GeV, are estimated
to be in the ranges of $(3.1\sim 4.9)\times 10^{-5}(F^{B\to \pi}_1(0)/0.33)^2$ 
and $(2.5\sim 4.2)
\times 10^{-5}(F^{B\to \pi}_1(0)/0.33)^2$ repesctively, 
with the combined branching ratio to
be $7.4\times 10^{-5}(F^{B\to \pi}_1(0)/0.33)^2$ 
almost independent of $\gamma$, which
are within the reach of B factories. $\bar B^0 \to \pi^+ X_q$ can
provide interesting information about the form factors. CP
violating asymmetries in these decays can be studied; and with the
current knowledge of the KM phase $\gamma$, we expect the asymmetries 
to be around 5\%.

\acknowledgments
This work was supported in part by the ROC NSC
under grant number NSC 90-2112-M-002-014, by the ROC Ministry of
Education Academic Excellence Project 89-N-FA01-1-4-3, 
and
by the U.S. D.O.E. under grant DE-FG-03-94ER40833.


\begin{references}
\bibitem{1}
H.-Y. Cheng and K.-C. Yang, Phys. Rev. {\bf D62}, 054029(2000);
G. Deshpande et al., Phys. Rev. Lett. {\bf 82}, 2240 (1999);
X.-G. He, W.-S. Hou and K.-C. Yang, Phys. Rev. Lett. {\bf 81}, 5738 (1998).
A. Ali, G. Kramer and C.-D. Lu, Phys. Rev. {\bf D58 }, 094009 (1998);
G. Deshpande, X.-G. He and J. Trameptic, Phys. Lett. {\bf B 345}, 547 (1995);
Y.-Y. Keum, H.-n. Li and A.I. Sanda, Phys. Rev. {\bf D63}, 054008(2001);
Phys. Lett. {\bf B504}, 6(2001);
C.-D. Lu, K. Ukai and M.-Z. Yang, Phys. Rev. {\bf D63}, 074009(2001);
D.-S. Du, D.-S. Yang and G.-H. Zhu, Phys. Lett. {\bf B 488}, 46 (2000);
Phys. Lett. {\bf B488}, 46(2000);
T. Muta, A. Sugamoto, M.-Z. Yang and Y.-D. Yang,
Phys. Rev. {\bf D 62}, 094020 (2000);
M.-Z. Yang and Y.-D. Yang, Phys. Rev. {\bf D 62}, 114019 (2000);
M. Beneke et al., Phys. Rev. Lett. {\bf 83}, 1914 (1999);
Nucl. Phys. {\bf B 591}, 313 (2000).

\bibitem{2} 
M. Neubert and J. Ronser, Phys. Lett. {\bf B441}, 403(1998);
Phys. Rev. Lett. {\bf 81}, 5074(1998);
M. Gronau and J. Rosner, Phys. Rev. {\bf D57}, 6843(1998);
X.-G. He et al., Phys. Rev. {\bf D64},
034002(2001); X.-G. He, C.-L. Hsueh and J.-Q. Shi, Phys. Rev. Lett.
{\bf 84}, 18(2000);
A. Buras and R. Fleischer, Eur. Phys. J. {\bf C16}, 97(2000).

\bibitem{3}
A. Datta et al., Phys. Rev. {\bf D 57}, 6829 (1998).

\bibitem{4}
X.-G, He, C.-H. Jin and J.-P. Ma, Phys. Rev. {\bf D64}, 014020(2001);
C.S. Kim et al., eprint hep-ph/0203093.

\bibitem{5}
N. G. Deshpande et al., Phys. Lett. {\bf B 366}, 300 (1996);
A. Kagan and A. Petrov, e-print hep-ph/9707354;
D. Atwood and A. Soni, Phys. Lett. {\bf B405}, 150(1997);
W.-S. Hou and B. Tseng, Phys. Rev. Lett. {\bf 80}, 434(1998);
D. Atwood and A. Soni, Phys. Rev. Lett. {\bf 81}, 3324 (1998);
A. Datta, X.-G. He and S. Pakvasa, Phys. Lett. {\bf B 419}, 369 (1998);
X.-G. He, J.P. Ma and C.-Y. Wu, Phys. Rev. {\bf D63}, 094004(2001);


\bibitem{6}
D. Ghosh et al., e-print hep-ph/0111106;
X.-G. He, J.-Y. Leou and J.-Q. Shi, Phys. Rev.
{\bf D64}, 094018(2001);
C.-K. Chua and W.-S. Hou, Phys. Rev. Lett. {\bf 86}, 2728(2001);
X.-G. He, W.-S. Hou and K.-C. Yang, Phys. Rev. Lett. {\bf 81}, 5738(1998);
M. Ciuchini, E. Gabrielli and G. Giudice, Phys. Lett. {\bf B388}, 353(1996);
F. Gabbiani et al., Nucl. Phys. {\bf B477}, 321(1996).

\bibitem{7}
D. Asner et al., CLEO collaboration, CONF 97-13, EPS332;
B. Aubert et al., BABAR collaboration, hep-ex/0109034;


\bibitem{8} K. Abe et al., e-print hep-ex/0201001.

\bibitem{9} G. Buchalla, A. Buras and M. Lautenbacher,
Rev. Mod. Phys. {\bf 68}, 1125 (1996);
A. Buras, M. Jamin and M. Lautenbacher, Nucl. Phys. {\bf B 400}, 75 (1993);
M. Ciuchini et al., Nucl. Phys. {\bf B 415}, 403 (1994).

\bibitem{10}
N. Deshpande and X.-G. He, Phys. Lett. {\bf B 336}, 471 (1994).


\bibitem{11} M. Wirbel, B. Stech and M. Bauer, Z. Phys. {\bf C29}, 637(1985);
M. Bauer, B. Stech and M. Wirbel, Z. Phys. {\bf 34}, 103(1987);
M. Bauer and B. Stech, Phys. Lett. {\bf B152}, 380(1985).

\bibitem{12} 
Patricia Ball, e-print hep-ph/9802394, JHEP 9809, 005(1998).

\end{references}
\end{document}